\documentstyle[aps,prb,multicol,graphicx,amstex]{revtex}
\title{A new broken $U(1)$-symmetry in extreme type-II superconductors}
\author{A. K. Nguyen$^{1}$ and A. Sudb{\o} $^{1,2}$}

\begin{document}
\maketitle
\begin{center}
  {\em $^1$ Norwegian University of Science and Technology, N-7034
    Trondheim, Norway \\ $^2$ California Institute of Technology,
    Pasadena CA91125, USA }
\end{center}

\begin{abstract}
A phase transition within the molten phase of the Abrikosov vortex
system without disorder in extreme type-II superconductors is found 
via large-scale Monte-Carlo simulations. It involves breaking a 
$U(1)$-symmetry, and has a zero-field counterpart, unlike vortex lattice 
melting. Its hallmark is the loss of number-conservation of connected
vortex paths threading the entire system {\it in any direction},
driving the vortex line tension to zero. This tension plays the 
role of a generalized ``stiffness'' of the vortex liquid, and serves
as a probe of the loss of order at the transition, where a weak specific 
heat anomaly is found. \\
Pacs-numbers: 74.20.De, 74.25.Dw, 74.25.Ha
\end{abstract}
\noindent

\begin{multicols}{2}
The key role of topological excitations in all phase-transitions,
whose proliferation is accompanied by the loss of a generalized
``stiffness'', has been emphasized by Anderson \cite{Anderson:Bo84}.
Recent numerical \cite{Nguyen:L96,Tachiki:L97,Nguyen:B98a,Nguyen:B98b}
and analytical work \cite{Tesanovic:B95,Tesanovic:cm98} reveals that
in extreme type-II superconductors, topological excitations in the
form of vortex loops are essential to the physics of the vortex system
in low magnetic fields.

It will be shown in this paper that the proliferation of large vortex
loops induces a phase transition {\it within the molten phase} of the
Abrikosov vortex system without disorder. The phase-transition thus
appears in a region of the phase-diagram where one normally would not
expect critical behavior simply based on a study of the {\it local
  superconducting order parameter $\Psi$} entering Ginzburg-Landau
theory. It is, however, a general property of extreme type-II
superconductors, distinct from the first order vortex lattice melting
transition in that it has a zero-field counterpart, i.e. the
transition from a superconductor to a normal metal mediated by a
vortex-loop ``blowout''
\cite{Dasgupta:L81,Williams:L87,Nguyen:B98a,Nguyen:B98b,Tesanovic:cm98}.

The model for extreme type-II superconductors considered in this paper, 
is the uniformly frustrated 3 dimensional anisotropic XY ($3DXY$) 
model \cite{Hetzel:L92},
defined by the Hamiltonian
\begin{equation}
  H(\{\theta({\bf r})\}) = - \sum_{{\bf r},\mu=x,y,z} J_\mu ~
        \cos[\nabla_\mu \theta({\bf r}) - A_\mu({\bf r}) ],
\label{Hamiltonian}
\end{equation}
where $\theta$ is the local phase angle of the superconducting complex
order parameter and $\nabla$ is a lattice derivative. The coupling
energy along the $\mu$-axis, $J_\mu$, is defined by $ J_x = J_y =
(\Phi_0^2 d)/(16 \pi^3 \lambda_{ab}^2) \equiv J_\perp$, and $J_z =
(\Phi_0^2 \xi_{ab}^2)/(16 \pi^3 \lambda_c^2 d)$, where $\lambda_{ab}$
and $\lambda_c$ are the magnetic penetration lengths set up by
screening currents in the CuO plane and along the crystal $c$-axis,
respectively.  $\Phi_0$ is the flux quantum, $\xi_{ab}$ is the
superconducting coherence length within the CuO-planes, and $d$ is the
distance between two $CuO$-layers in {\em adjacent unit cells}. We may
use the lattice spacing as a measure of $\xi_{ab}$. In Eq.
\ref{Hamiltonian}, $A_\mu$ is related to the non-fluctuating external
magnetic vector potential ${\bf A}_{vp}$ by $A_\mu({\bf r}) \equiv
2\pi/\Phi_0 \int_{{\bf r}}^{{\bf r} + \hat{e}_\mu} d{\bf r}' \cdot
{\bf A}_{vp}({\bf r}')$, where $\hat{e}_\mu$ is the unit vector along
the $\mu$-axis.

We consider systems of size $L_x \times L_y \times L_z$. To perform
simulations and finite size scaling of systems with very low filling
fractions, we define an ``extended'' Landau gauge
\begin{eqnarray}
A_x=\frac{2 ~\pi ~y ~ m_y ~n ~ m}{L_x L_y};~~
A_y=\frac{2~ \pi ~x ~ n_x ~m ~ n}{L_x L_y},  
\end{eqnarray}
where $n_x,n,m_y,m$ are positive integers satisfying $n_x ~ n =  L_y$, and 
$m_y  ~ m = L_x$. Hence, the filling fraction $f$ is given by
$f = n ~ m ~ [n_x-m_y]/L_x L_y$.

In this paper, we consider the filling fractions $f$ in the range $1/f
\in (20,..,1560)$.  {\it These filling fractions are so low that
  spurious pinning to the numerical lattice is eliminated, i.e. any
  spurious transverse Meissner-effect has vanished, well below all
  temperatures of interest in this paper}
\cite{Nguyen:L96,Tachiki:L97,Nguyen:B98a,Nguyen:B98b}.  Using $f = B
\xi^2_{ab}/\Phi_0$ and $\xi_{ab} = 15 \AA$, $1/f = 1560$ corresponds
to a uniform induction $B \sim 0.56 T$. Using $\lambda_{ab} > a_v$ as
a condition for uniform $B$ and $\lambda_{ab} = 1500 \AA$, we find
that this assumption is valid for inductions as low as $0.1 T$,
corresponding to a filling fraction of order $10^{-4}$. The uniform
induction $B$ is taken along the crystal $\hat{c}$-axis. The
anisotropy parameter is given by $\Gamma = \sqrt{J_\perp/J_z} =
\lambda_cd/\lambda_{ab}\xi_{ab}$. For each filling fraction, we carry
out simulations on three systems with different sizes ($\sim 40^3,
\sim 80^3, \sim 150^3$) to study finite-size effects. For the specific
heat, we also consider $360^3$ for $f=1/90$.

In addition to the specific heat $C$ \cite{Nguyen:B98a,Nguyen:B98b},
we calculate three quantitities in this paper. The first two are
standard, while the third is unusual and probes a subtle
$U(1)$-symmetry of the system.

{\underline{\it{ i) Helicity modulus}}}. To probe the global phase
coherence we consider the helicity moduli, $\Upsilon_{\mu}$, defined
as the second derivative of the free energy with respect to an applied
phase twist \cite{Li:B93,Nguyen:B98b}.  When $\Upsilon_\mu$ is zero,
the resistance along $\mu$-direction is finite, and any applied
current along $\mu$-direction will dissipate energy. Note that
$\Upsilon_\mu$ is a global quantity, so even when $\Upsilon_\mu=0$ the
system can still maintain local phase coherence and exhibit
diamagnetic fluctuations. $\Upsilon_z$ will vanish at the temperature
$T_z$, coinciding with the zero-field superconducting transition $T_c$
when $f=0$, and will coincide with the melting temperature of the
Abrikosov vortex lattice (AVL) when $f > 0$.

{\underline{\it{ ii) Structure factor}}}. To probe the Abrikosov
vortex lattice (AVL) melting, we consider the structure function for
$q_z$ vortex segments \cite{Cavalcanti:E92,Nguyen:L96}.  A vortex
configuration can be obtained from a phase-configuration from the
counterclockwise line integral of the gauge-invariant
phase-differences around any plaquette of the numerical lattice. It
must satisfy the condition for conservation of vorticity,
$\sum_{\cal{C}} [\nabla_\nu \theta({\bf r}) - A_\nu({\bf r})] =
2\pi(q_\mu({\bf r}) - f_\mu)$.  Here, $\cal{C}$ is the closed path
traced out by the links surrounding an elementary plaquette, and $\nu$
represents the Cartesian components of the links comprising the closed
path $\cal{C}$. Furthermore, $q_\mu({\bf r})=0,\pm 1$ represents a
vortex segment penetrating the plaquette enclosed by $\cal{C}$. If we
monitor the structure function $S({\bf Q},k_z=0)$ of one chosen Bragg
peak as a function of temperature, we expect that $S({\bf Q},k_z=0)$
has a discontinuous drop to a very small value at the melting
temperature. The structure function will vanish discontinuously at the
melting temperature $T_m(B)$ of the AVL.

{\underline{\it{ iii) Vortex-path probability $O_L$}}}. We define
$O_L$ as the probability of finding a directed vortex path threading
the entire system transverse to the induction $B$, {\em without using
  the PBC along the field direction}. It is obtained by computing the
number $N_V$ of times we find {\it at least one} such path threading
the system in any direction $\perp B$ in $N_P$ different
phase-configurations, normalized by $N_{P}$, i.e. $O_L=N_V/N_P$.  Note
that this differs from algorithms employed in previous publications
\cite{Jagla:B96,Koshelev:B97}.

$O_L=0$ means that there is no connected path of vortex segments that
threads the entire system in the transverse direction, without using
PBC along the field direction several times. Now, let $N^{\alpha}_L$
($\alpha \in [x,y,z]$) denote the areal density of connected vortex
paths threading the system in any direction, including the direction
parallel to the induction. It is clear that in the AVL phase $O_L=0$,
and $N^z_L=B/\Phi_0$, while $N^x_L = N^y_L=0$. Thus, $N_L^{\alpha}$ is
a conserved quantity at fixed induction $B$.  On the other hand,
$O_L=1$ implies that $N^{x,y}_L > 0$, and the {\it total} number of
vortex paths threading the system in any direction scales with system
size, but undergoes thermal fluctuations. Therefore, $N_L^{\alpha}$ is
no longer a conserved quantity. The change in $O_L$ takes place at the
temperature $T_L(B)$, which coincides with $T_c$ when $f=0$.

Number conservation uniquely identifies a $U(1)$-symmetry, and hence
the low-temperature phase of the {\it vortex}-system (the dual of the
phase-representation of the superconductor) exhibits explicit
$U(1)$-symmetry, since $O_L=0$. At high temperatures $O_L=1$,
$N^{\alpha}_L$ is not conserved, and the $U(1)$-symmetry is broken.
Under no circumstance can a $U(1)$-symmetric phase be analytically
continued to a $U(1)$-nonsymmetric one. {\it The change in $O_L$ from
  $0$ to $1$ therefore signals a geometric transition taking place
  withing the vortex-liquid}, indicative of the breakdown of
vortex-{\it line} liquid picture of the molten phase of the Abrikosov
vortex lattice.

\begin{figure}
  \begin{picture}(0,205)(0,0)
     \put(-60,-200)
         {\includegraphics[angle=0,scale=0.55]
         {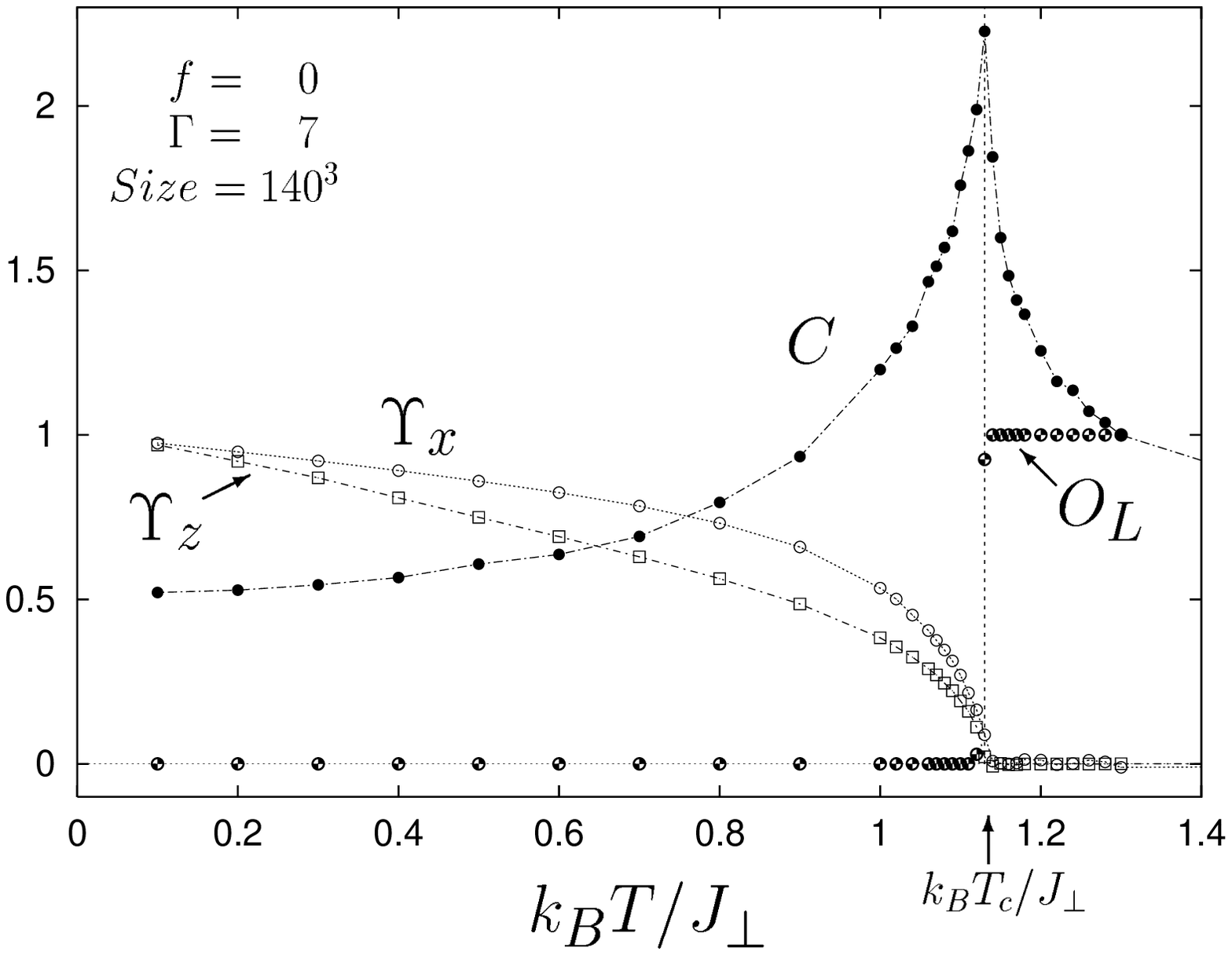}}
  \end{picture}
{\small FIG. \ref{Zero.Field}. 
Specific heat, vortex-path probability $O_L$, and 
helicity moduli along $x$-direction ($\Upsilon_x$) and along $z$-direction 
($\Upsilon_z$) as functions of temperature for the zero field case. 
$\Gamma = 7$, and system size is $140^3$. For $T < T_c$,  $O_L = 0$. 
For $T > T_c$ a vortex loop blow out has taken place,
i.e. $O_L = 1$.}
\refstepcounter{figure}
\label{Zero.Field}
\end{figure}

We first discuss the zero-field simulation results. In Fig.
\ref{Zero.Field}, we plot the specific heat $C$, the helicity moduli
$\Upsilon_x$,$\Upsilon_z$, and $O_L$ as functions of temperature for a
system with $Size = 140^3$ and anisotropy $\Gamma=7$. $C$ has a
near-logarithmic singularity at $k_BT_c/J_\perp = 1.12$, where also
$\Upsilon_x$ and $\Upsilon_z$ drop to zero.  This second order phase
transition is the superconducting-normal state phase transition. Note
that both $\Upsilon_z$ and $\Upsilon_x$ vanish at $T_c$, albeit with
different amplitudes due to the anisotropy of the model. Since there
is one, and only one, transition in zero field, this is a check that
the system sizes used in the simulations are adequate for the given
anisotropy \cite{Nguyen:B98a}.

\vspace{-0.3cm}

For $T < T_c$, it is clear that $O_L=0$ and $N^\alpha_L$ is conserved
and equal to zero. If for $T > T_c$, we have $O_L=1$, then
$N^{\alpha}_L >0$.  This feature of $O_L$ is precisely what we find in
our simulations, see Fig. \ref{Zero.Field}. This means that the system
undergoes a $3DXY$-transition, with a low-temperature $U(1)$-symmetric
state and a high-temperature $U(1)$-nonsymmetric state. Note that in
terms of the ordinary superconducting orderparameter $\Psi$ in
Ginzburg-Landau theory in zero field, the situation is the opposite as
far as $U(1)$-symmetry and symmetry breaking is concerned: The more
familiar phase-representation of the superconductor is related to the
vortex-representation via a duality transformation which interchanges
low and high temperatures \cite{Kleinert:Bo89}.

\begin{figure}
  \begin{picture}(0,320)(0,0)
     \put(-30,-20)
         {\includegraphics[angle=0,scale=0.45]
         {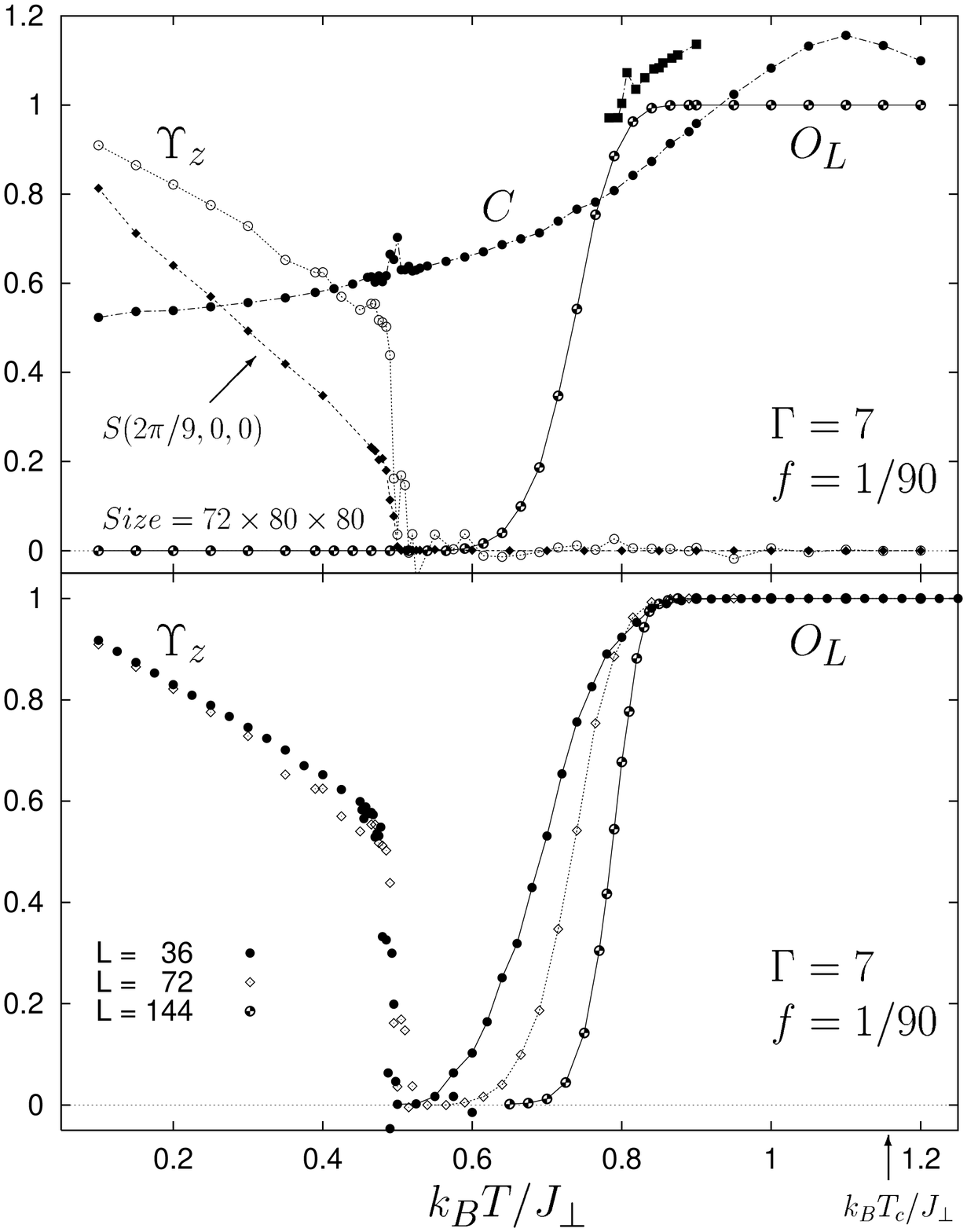}}
  \end{picture}
  {\small FIG. \ref{Finite.Field}. Top panel: Specific heat $C$,
    structure factor $S$, vortex-path probability $O_L$ and helicity
    moduli $\Upsilon_x$, $\Upsilon_z$ as functions of temperature for
    a system with a filling fraction $f=1/90$ and $Size=72 \times 80
    \times 80$. The melting temperature is given by $k_BT_m/J_\perp =
    0.49$. At $T_L(B)$, we also show $C$ for a system of size $360^3$,
    illustrated by solid squares.  The curve is shifted up by the
    amount $0.2 ~ k_B$ from the curve for i $72 \times 80 \times 80$,
    for clarity.  {\it Note the increase in the weak anomaly at
      $T_L(B)$}, which is the hallmark of a thermodynamic
    phase-transition. \\  
    Bottom panel: Helicity modulus along the field direction
    $\Upsilon_z$ and $O_L$ as functions of temperature for three
    systems with $f=1/90$ and sizes $36 \times 40 \times 40$, $72
    \times 80 \times 80$, $144 \times 160 \times 160$. }
\label{Finite.Field}
\end{figure}

Next, we consider finite magnetic fields. In Fig. \ref{Finite.Field}
top panel, we plot the specific heat $C$, the structure factor $S$,
$O_L$ and the helicity moduli $\Upsilon_x$ \cite{Upsilonx} and
$\Upsilon_z$ as functions of temperature for a system with filling
fraction $f=1/90$, $Size=72 \times 80 \times 80$ and $\Gamma=7$. We
see from the top panel of Fig. \ref{Finite.Field} that at
$k_BT_m/J_\perp = 0.49$ the structure factor $S$ drops discontinuously
from $0.2$ to $0$, indicating a first order AVL melting transition at
$T_m$. At $T=T_z$, $\Upsilon_z$ drops discontinuously from $0.55$ to
$0$. We see that in our simulations $T_z=T_m$, indicating that the AVL
melts directly into an incoherent vortex liquid. Thus, in the vortex
liquid phase there is no global phase coherence in any direction
\cite{Nguyen:B98a,Tachiki:L97,Nguyen:B98b}.  This conclusion also
holds for the isotropic case $\Gamma = 1$ \cite{Nguyen:B98c,Ryu:cm98}.

In the top panel of Fig. \ref{Finite.Field}, the specific heat shows a
spike precisely at $T_m$, indicating a discontinuous jump in the
internal energy, and thus proving the existence of a first order phase
AVL melting transition with a latent heat at $T_m$. We now focus on
the temperature range where $O_L$ changes from $0$ to $1$. For the
filling fraction $f=1/90$, $O_L$ rises from $0$ to $1$, and reaches
$1$ at $T = T_L$ well separated from $T_m$ and the crossover
temperature $T_{Bc2}$, see top panel of Fig. \ref{Finite.Field}.
vortex loop blowout takes place. For increasing system size the width
of the transition decreases rapidly, see the lower panel of Fig.
\ref{Finite.Field}.  {\it Note also how the low-temperature tail of
  $O_L$ is suppressed when increasing the system size, while the
  temperature $T_L(B)$ where $O_L$ reaches the value $1$ stays almost
  fixed, moving slightly down with increasing system size.} Thus, in
the thermodynamic limit, there exists a well defined temperature $T_L$
where $O_L$ rises sharply from $0$ to $1$. This is found for all the
filling fractions considered. {\it For $f=1/90$, we have also
  performed simulations for systems of size $360^3$, to bring out the
  scaling of the additional weak specific heat anomaly at $T_L(B)$,
  see the top panel of Fig. \ref{Finite.Field}.} It shows conclusively
that $T_L(B)$ is a thermodynamic phase-transition.

The position of the specific heat anomaly at $T_L(B)$ for the system
of size $360^3$ appears at a slightly lower temperature than those
where $O_L$ starts to dip down from $1$ in smaller systems. This is
because the $O_L$-curves become sharper as $L$ increases, and the
position of the $C$-anomaly is at a position which may be estimated to
be the limiting temperature at which $O_L$ changes abruptly from $0$
to $1$. Note that in the specific heat, the anomalies both at $T_m(B)$
and $T_L(B)$, both equal in magnitude, are well beyond the noise-level
in the simulations.  For the other parts of the $C$-curve, the
uncertainties are of order the symbol size.

The fact that $O_L=1$ above $T_L(B)$, indicates that the {\it
  long-wavelength } vortex-line tension $\varepsilon(T)$ vanishes. We
may infer this from the change in $O_L$.

Furthermore, we have checked how $T_L(B)$ varies with a change of
aspect ratio $L_x/L_z$ of the system. For $f=1/380$, $\Gamma=7$, (not
shown) we have considered systems of size $L_x \times L_y \times
L_z/\alpha$, with $L_x=L_y=L_z=40,80,120$, varying $\alpha \in
[1.00,1.25,1.50,1.75,2.00]$.  Within a vortex-line liquid picture, we
should find $T_L \propto \alpha$ \cite{Nordborg:PC}. Instead, we find
a change of less than $5 \%$ on increasing $\alpha$ from $1$ to $2$,
again difficult to explain within a vortex-line liquid picture of the
molten phase above $T_L(B)$.  This is supportive of a change in the
connectivity of the vortex-tangle at $T_L(B)$, in a direction
perpendicular to the magnetic field. In a vortex liquid with non-zero
$\varepsilon(T)$, the vortex system is only connected across the
system along the field direction.

The long-wavelength vortex-line tension may thus serve as a probe of
the loss of order at the transition, playing the role of a generalized
``stiffness'' \cite{Anderson:Bo84} characterizing the two vortex
liquid phases above and below $T_L(B)$. More precisely, the two
vortex-liquid regimes are characterized by breaking a $U(1)$-symmetry
of the {\it dual } theory of the Ginzburg-Landau theory, on crossing
the line $T_L(B)$ from below.

\begin{figure}
  \begin{picture}(0,210)(0,0)
     \put(-20,-105)
         {\includegraphics[angle=0,scale=0.42]
         {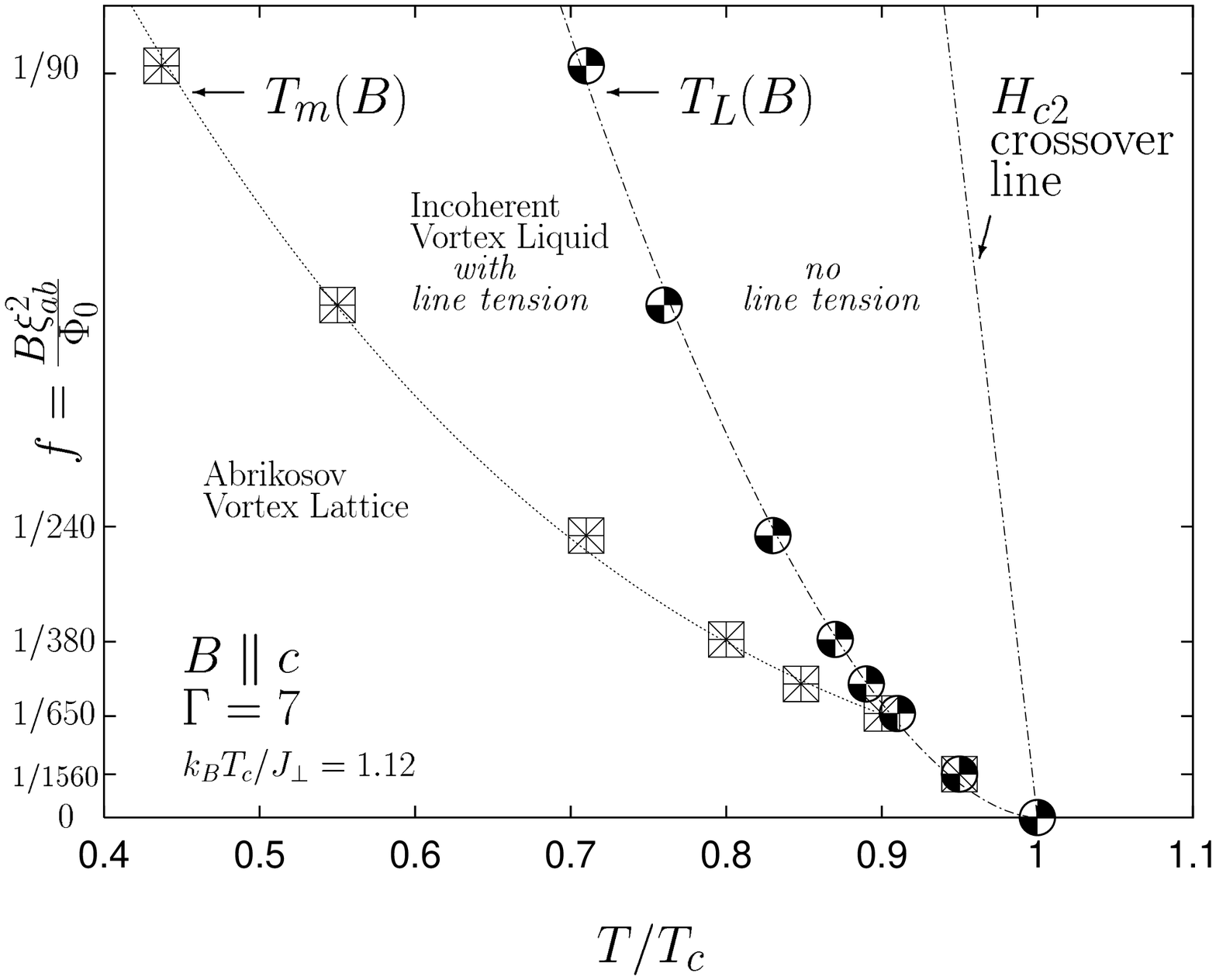}}
  \end{picture}
  {\small FIG. \ref{Phase.Diagram}. The intrinsic $(B,T)$-phase
    diagram of an extreme type-II superconductor, obtained from the
    uniformly frustrated 3DXY model with $\Gamma = 7$ and $B \parallel
    c$. $k_BT_c/J_\perp = 1.12$. The filling fractions $f$ that are
    used are given by $1/f = 90,132,290,380,506,650,1560$. The lines
    are guides to the eye.  The $H_{c2}$-line is found from the
    maximum value of the broadened peak of the specific heat in a
    finite magnetic field.}
\label{Phase.Diagram}
\end{figure}

We have obtained $C,~\Upsilon_z,~S$, and $O_L$ for a wide range of
filling fractions, see Fig. \ref{Phase.Diagram}. The resulting
$(B,T)$-phase diagram for the uniformly frustrated 3DXY model with the
anisotropy parameter $\Gamma = 7$ and $B \parallel c$, is shown in
Fig. \ref{Phase.Diagram}.  In zero field, a $3DXY$-transition
separates the superconducting phase from the normal phase. In a finite
magnetic field, we have three distinct phases: the AVL for $T <
T_m(B)$, the vortex liquid with line tension for $T_m(B) < T <
T_L(B)$, and the vortex liquid without line tension for $T > T_L(B)$.
{\em $T_L(B)$ is a critical line, the finite field counter part of the
  zero field vortex loop blowout at $T=T_c$}. In the low field regime,
$f < 1/600$, the AVL melts directly into a vortex liquid with no
linetension.

The dimensionless criterion determining the low-field regime of the
melting line $T_m(B)$, is that $f \Gamma^2 << 1$ \cite{Nordborg:B99}.
For our case, the merging of $T_m(B)$ and $T_L(B)$ occurs at $f
\Gamma^2 \approx 1/13$, which is well within the low-field regime.
Note that at more elevated fields, $f \Gamma^2 > 1/4$, the vortex
lattice melting line is well described by both the $XY$-model and the
$2D$ boson-analogy of the vortex-system, as recently nicely elaborated
on by Koshelev and Nordborg \cite{Nordborg:B99}. However, from their
Fig. 1, we note that {\it deviations} from universal properties of
{\it linelike} melting of the vortex lattice starts to appear for $f
\Gamma^2 < 1/5$, consistent with the above picture.

$T_m(B)$ meets $T_L(B)$ at a tricritical point, see Fig.
\ref{Phase.Diagram}.  {\it The continuation of the melting line below
  the tricritical point is first order}. We have observed a
substantial increase in the $\delta$-function spike at the melting
transition below the tricritical point, since the entropy in the
transition at $T_L(B)$ now contributes to a first order transition
\cite{Nguyen:B98c}.  Note also that well above the tricritical point,
the position of $T_m(B)$ is adequately described by a
Lindemann-criterion applied to the London-model including only
field-induced vortex lines \cite{Houghton:B89}. It is not clear,
however, that such a model would give $\Upsilon_z = 0$ in the entire
molten phase.

In summary, we have used large-scale Monte-Carlo simulations to
analyze the characteristics of vortex-paths in a model of extreme
type-II superconductors in the absence of disorder. An abrupt change
in this characteristics reveals a phase-transition involving the
breaking of a $U(1)$-symmetry, which is suggested to exist in zero as
well as finite magnetic field. It is therefore distinct from the
vortex lattice melting transition. This symmetry breaking is a
consequence of a thermally driven vortex-loop ``blowout''
\cite{FFH:B91}. It results in a nonconserved number of thermally
fluctuating connected vortex paths threading the entire system in any
direction, including the direction perpendicular to a transverse
magnetic field.  {\it In other words, the connectivity of the
  vortex-tangle changes \cite{Jagla:B96}, and the insensitivity of
  $T_L(B)$ strongly suggests that this is a feature that survives in
  the thermodynamic limit}.

In principle it is also possible for a $U(1)$-symmetry breaking to be
first-order, although in zero magnetic field this only happens for
small values of $\kappa$ \cite{Halperin:L74}. In finite field this
could change. More work is clearly needed to investigate in detail the
universality class of the proposed transition at $T_L(B)$.

\begin{center}
  ***
\end{center}

Support from the Research Council of Norway under Grants No.
110566/410, No. 110569/410, and a grant for computing time under the
Program for Supercomputing, is gratefully acknowledged. We thank L.
Bulaevskii, S.-K. Chin, M. Dodgson, {\O}. Fischer, A. Koshelev, J. M.
Kosterlitz, M. Moore, and T. Natterman for discussions. In particular,
we thank Z. Te{\v s}anovi{\'c} for stimulating discussions and
comments, and J. Amundsen for his invaluable assistance in optimizing
our computer codes for use on the CrayT3E.

\end{multicols}
\end{document}